\documentclass[useAMS,usenatbib]{mnras}

\usepackage[english]{babel}
\usepackage{graphicx,booktabs}
\usepackage{blindtext}
\usepackage{xcolor}
\usepackage{amssymb,amsmath, wasysym}
\usepackage{tikz}
\usetikzlibrary{backgrounds}
\usetikzlibrary{arrows}
\usetikzlibrary{shapes}

\usepackage{xspace}
\usepackage{siunitx}

\def\apj{ApJ}%
\def\apjl{ApJ}%
\def\apjs{ApJS}%
\def\ao{Appl.~Opt.}%
\def\aap{A\&A}%
\def\mnras{MNRAS}%
\def\prd{Phys.~Rev.~D}%
\def\prl{Phys.~Rev.~Lett.}%
\def\nat{Nature}%
\def\physrep{Phys.~Rep.}%
\def\jcap{JCAP}

\newcommand{\lcdm}{$\Lambda$CDM\xspace}

\newcommand{\mn}{{\mu\nu}}

\newcommand{\fr}{$f(R)$-gravity\xspace}
\newcommand{\fR}{f_{\rm R}}

\newcommand{\msh}{\, h^{-1}{\rm M}_\odot}
\newcommand{\hompc}{\,h {\rm Mpc}^{-1}}

\newcommand{\mpcoh}{\,h^{-1} {\rm Mpc}}

\begin{document}

\title[Galaxy formation in $f(R)$ gravity: Matter statistics]
{Simulating galaxy formation in $f(R)$ modified gravity:\\
Matter, halo, and galaxy-statistics}

\author[C. Arnold \& B. Li]
{Christian Arnold,\hspace{-.25em}$^1$\thanks{E-mail: christian.arnold@durham.ac.uk}
Baojiu Li$^1$
\\$^{1}$Institute for Computational Cosmology, Department of Physics, Durham University, South Road, Durham DH1 3LE, UK
}

\date{\today}
\maketitle

\begin{abstract}
We present an analysis of the matter, halo and galaxy clustering in \fr employing the SHYBONE full-physics hydrodynamical simulation suite. Our analysis focuses on the interplay between baryonic feedback and \fr in the matter power spectrum, the matter and halo correlation functions, the halo and galaxy-host-halo mass function, the subhalo and satellite-galaxy count and the correlation function of the stars in our simulations.
Our studies of the matter power spectrum in full physics simulations in \fr show, that it will be very difficult to derive accurate fitting formulae for the power spectrum enhancement in \fr which include baryonic effects. 
 We find that the enhancement of the halo mass function due to \fr and its suppression due to feedback effects do not show significant back-reaction effects and can thus be estimated from independent GR-hydro and $f(R)$ dark matter only simulations. Our simulations furthermore show, that the number of subhaloes and satellite-galaxies per halo is not significantly affected by \fr. Low mass haloes are nevertheless more likely to be populated by galaxies in \fr. This suppresses the clustering of stars and the galaxy correlation function in the theory compared to standard cosmology.

\end{abstract}


\renewcommand{\d}{{\rm d}}

\section{Introduction}
\label{sec:introduction}

The standard theory of gravity, \textit{Einstein's General Relativity} (GR), is confirmed to remarkably high precision in our local environment and other small scale systems \citep{will2014}. Together with the cosmological constant $\Lambda$ and cold dark matter (DM), it is the key ingredient for the current standard model of cosmology, the \lcdm model. The \lcdm model provides a very successful description of the large scale structure in and the expansion history of our universe \citep{joyce2015}. 

Beyond these local tests and binary pulsars there are nevertheless very little test of gravity on intermediate and large scales. Upcoming large scale structure surveys such as EUCLID \citep{euclid} or LSST \citep{lsst} aim to test gravity on these large scales. In order to do so they require a detailed understanding of how possible deviations from GR and \lcdm cosmology alter the large scale structure. Such an understanding can be provided by cosmological simulations of alternative models of gravity like the ones we present and analyse in this work. 
Possible alternatives to GR which were previously studied in cosmological simulations include the galileon model \citep{li2013}, the symmetron model \citep{llinares2014}, the nDGP model (normal branch Dvali, Gabadadze, and Porrati gravity, \citealt{schmidt2009c, li2013c}) and \fr which we consider in this work.

\fr is a generalization of GR which can be used as a model to predict in which cosmological and astrophysical observables possible deviations from the standard theory of gravity would be testable and in which way the data provides by upcoming surveys should be analyzed in order to test gravity. 
Among its GR-testing abilities, \fr has several advantages \citep[see e.g.][for a recent review]{joyce2015}. The theory features the chameleon  screening mechanism \citep{khoury2004} which ensures that the modifications to gravity are screened in high density environments like the solar system. In unscreened regions, the gravitational force is enhanced by a factor of $4/3$. The non-linearity of the equations introduced by the chameleon mechanism limits the applicability of perturbative methods and makes cosmological simulations the most successful tool to study structure formation in the theory. \fr furthermore predicts a speed of gravitational waves which is equal to the speed of light and is therefore consistent with the results of \cite{GW2017}. The theory is also theoretically very well understood. 

On intermediate and astrophysical scales the effects of modified gravity are nevertheless often degenerate with the influences of astrophysical processes which are primarily driven by baryonic feedback. It has therefore been claimed that DM-only simulations in modified gravity are not sufficient to find reliable astrophysical tests of gravity but that baryonic physics has to be simulated at the same time \citep{puchwein2013, arnold2018}. 
In this work we analyse the SHYBONE cosmological simulation suite \citep{arnold2019} which combines a full-physics description of baryonic processes using the Illustris TNG model \citep{pillepich2018b, springel2018, genel2018, marinacci2018, nelson2018} and a fully non-linear description of \cite{husa2007} \fr in the Newtonian limit. 

The simulation suite includes full-physics simulations in \fr and \lcdm cosmology featuring descriptions for hydrodynamics on a moving mesh, star formation and feedback, galactic winds, feddback from active galactic nuclei (AGN) and magnetic fields \citep{pillepich2018, weinberger2017}. They are complemented by DM-only simulations using the same initial conditions and also non-radiative hydrodynamical simulations which use a very basic description of hydrodynamics ignoring feedback processes. 

The simulations used in this work are the only simulations which include a complete hydrodynamical model and $f(R)$ modified gravity at the same time to date. Alongside the \textsc{arepo} code which was used for these simulations, several other cosmological simulations codes for \fr exist \citep[see e.g.][]{oyaizu2008,li2012, puchwein2013, llinares2014}. Previous works on simulations in \fr include studies of the matter and halo distribution \citep{schmidt2010, zhao2011, li2011, lombriser2013, puchwein2013, hellwing2013, hellwing2014, arnold2015, cataneo2016, arnalte2017}, void properties \citep{zivick2015, cautun2018}, the velocity dispersion of DM haloes \citep{schmidt2010, lam2012, lombriser2012b}, cluster properties \citep{lombriser2012, arnold2014, he2016, mitchell2018, mitchell2019}, weak gravitational lensing \citep{shirasaki2015, li2018} and redshift space distortions \citep{jennings2012}. Non-radiative hydrodynamical simulations have been used to study galaxy clusters and the Lyman-$\alpha$ forest in \fr \citep[e.g.][]{arnold2014, arnold2015, hammami2015}. 

This work is structured as follows. In Section \ref{sec:gravity} we then introduce the theory of \fr. We introduce the simulation suite and give a brief overview over the Illustris TNG galaxy formation model in Section \ref{sec:methods}. In Section \ref{sec:results} we finally present our results which we conclude and discuss in Section \ref{sec:conclusions}. 

\begin{table*}
\centering
\begin{tabular}{l l l c c c c c}
\toprule
Simulation & Hydro model & Cosmologies & $L_{\rm box} [\mpcoh]$ & $N_{\rm DM}$ & $N_{\rm gas}$ & ${m}_{\rm DM} [\msh]$ & $\bar{m}_{\rm gas} [\msh]$ \\
\midrule
Full-physics, large box & TNG-model & \lcdm, F6, F5 & $62$ & $512^3$ & $\approx 512^3$ &{$1.3 \times 10^8 $} &  {$\approx 3.1 \times 10^7$}\\
Full-physics, small box & TNG-model & \lcdm, F6, F5 & $25$ & $512^3$ & $\approx 512^3$ &{$8.4 \times 10^6$} &  {$\approx 2.2 \times 10^6$}\\
Non-rad & Non-radiative & \lcdm, F6, F5 & $62$ & $512^3$ & $\approx 512^3$ & {$1.3 \times 10^8 $} &  {$\approx 3.6 \times 10^7$}\\
DM-only & -- & \lcdm, F6, F5, F4 & $62$ & $512^3$ & -- & $1.5 \times 10^8 $ &  --\\
\bottomrule
\end{tabular}
\caption{An overview over the SHIBONE simulation suite.}
\label{tab:sims}
\end{table*}

\section{$f(R)$-gravity}
\label{sec:gravity}
\fr \citep{buchdahl1970} is a generalisation of Einstein's general relativity (GR). It introduces an additional scalar degree of freedom which leads to a so called fifth force, enhancing gravity by $4/3$ in low density environments while deep gravitational potentials are screened from the fifth force and thus experience standard GR-like gravity. 
The theory is constructed in the following way: Using the same framework as GR one adds a scalar function $f(R)$ of the Ricci scalar $R$ to the action:
\begin{align}
S=\int {\rm d}^4x\, \sqrt{-g} \left[ \frac{R+f(R)}{16\pi G} +\mathcal{L}_m \right],\label{action}
\end{align}
where $G$ is the gravitational constant, $g$ is the determinant of the metric $g_\mn$ and $\mathcal{L}_m$ is the Lagrangian of the matter fields. Varying the action with respect to the metric, leads to the field equations of (metric) \fr:

\begin{align} G_{\mu\nu} + \fR R_{\mu\nu}-\left( \frac{f}{2}-\Box \fR\right) g_{\mu\nu} - \nabla_\mu \nabla_\nu \fR = 8\pi G T_{\mu\nu} \label{Eequn},  \end{align} 

where $G_\mn$ and $R_\mn$ denote the components of the Einstein and Ricci tensor, respectively. The scalar degree of freedom, $\fR$, is the derivative of the scalar function $\fR \equiv \d f(R)/ \d R$. The energy momentum tensor is $T_\mn$, covariant derivatives are written as $\nabla_\nu$ and $
\Box \equiv \nabla_\nu \nabla^\nu$, where Einstein summation is used.

In the context of cosmological simulations one commonly works in the weak-field, quasi-static limit \citep[a discussion on the validity of this assumption for \fr can be found in][]{sawicki2015} in which the above equations (\ref{Eequn}) simplify considerably. In this limit, the gravitational potential $\Phi$ is given by a modified Poisson equation
\begin{align}
 \nabla^2 \Phi = \frac{16\pi G}{3}\delta\rho - \frac{1}{6} \delta R,\label{poisson}
\end{align}
where $\delta R = R - \bar{R}(a)$ depends on the scalar degree of freedom, which is governed by a second differential equation
\begin{align}
 \nabla^2 \fR =  \frac{1}{3}\left(\delta R -8\pi G\delta\rho\right). \label{fRequn}
\end{align}

Before equations (\ref{poisson}) and (\ref{fRequn}) can used to calculate the gravitational potential, they have to be connected by choosing a functional form for $f(R)$. Here, we adopt the widely studied model proposed by \cite{husa2007}:
\begin{align}
 f(R) = -m^2\frac{c_1\left(\frac{R}{m^2}\right)^n}{c_2\left(\frac{R}{m^2}\right)^n +1},\label{fr}
\end{align}
where $m^2 \equiv \Omega_m H_0^2$. We choose $n=1$. $c_1$ and $c_2$ are two additional parameters.
 For an appropriate choice of parameters this model passes the very stringent constraints constraints on gravity within the solar system \citep{will2014} by screening the modifications to gravity in high density environments through the chameleon mechanism. 
The model also features a cosmic expansion history which is very close to that of a \lcdm universe if one chooses \citep{husa2007}
\begin{align}
\frac{c_1}{c_{\color{magenta}2}} = 6 \frac{\Omega_\Lambda}{\Omega_m} && \text{and} && \frac{c_2\, R}{m^2} \gg 1.
\end{align}
Within this framework, one can approximate the scalar degree of freedom, $\fR$,  as 
\begin{align}
\fR \equiv \frac{\d f(R)}{\d R} = -n\frac{c_1\left(\frac{R}{m^2}\right)^{n-1}}{\left[c_2\left(\frac{R}{m^2}\right)^n+1\right]^2}\approx-n\frac{c_1}{c_2^2}\left(\frac{m^2}{R}\right)^{n+1}.\label{fR}
\end{align}

Adopting the Friedmann-Robertson-Walker metric the remaining free parameter can be expressed in terms of $\bar{f}_{R0}$ which is the background value of the scalar degree of freedom at redshift $z = 0$. This parameter sets the onset threshold for chameleon screening (in terms of gravitational potential depth). In this work, we consider four different values for $\bar{f}_{R0}$ (and GR): The F6 model ($\bar{f}_{R0} = -10^{-6}$) screens objects already at relatively low gravitational potential depth and is therefore consistent with most observational constraints \citep{terukina2014}. The F5 model ($\bar{f}_{R0} = -10^{-5}$) only screens regions with greater potential depth and therefore features more prominent MG effects but is in tension with solar system constraints. Finally we also consider the F4 model ($\bar{f}_{R0} = -10^{-4}$), which shows very strong $f(R)$-effects but fails most observational tests. As both the F5 and the F4 model are ruled out by observational data, they serve as toy models which help to better understand the general behaviour of chameleon screening models.


\begin{figure*}
\centering
\includegraphics[width = \textwidth]{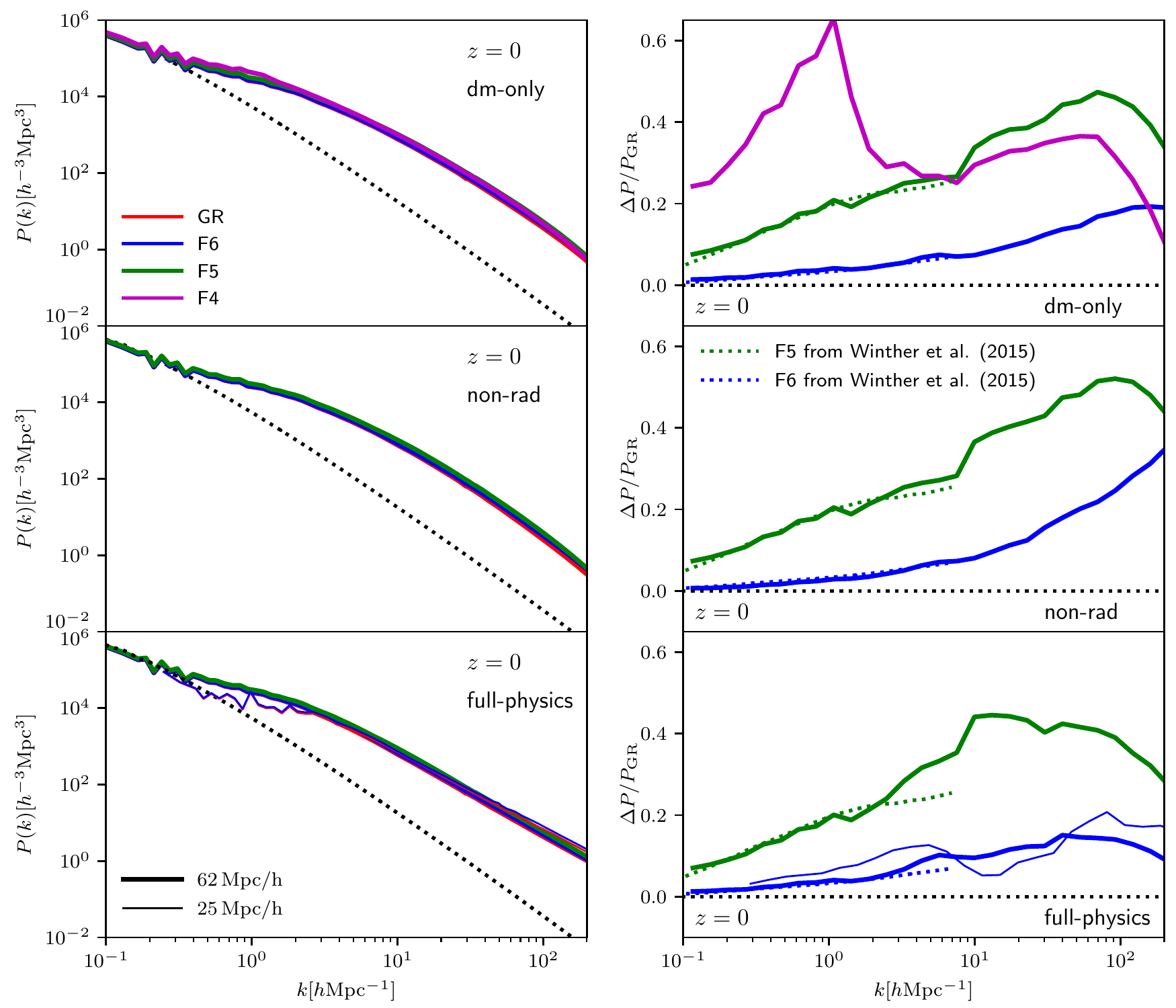}
\caption{The 3D matter power spectrum for the 9 different runs of our simulation suite at redshift $z=0$. Results from the collisionless DM-only simulations are show in the \textit{top panels}, from the non-radiative simulations in the \textit{middle panels} and from the full-physics hydrodynamical runs in the \textit{bottom panels}. 
The \lcdm results are displayed in \textit{red}, results for F6, F5 and F4 (only for DM-only) are shown in \textit{blue}, \textit{green} and \textit{magenta}, respectively. 
Thick solid lines show results from the $62\mpcoh$ simulations boxes, thin solid lines results from the $25\mpcoh$ simulation boxes (note that the $25\mpcoh$ results are only shown for full-physics; the small box F5 simulation was only run until $z=1$).
The \textit{left panels} show the absolute value of the power spectra, the \textit{right panels} the relative difference of the \fr results to the corresponding standard gravity run. We show the DM-only simulation results from the modified gravity code comparison project \protect\citep{winther2015} as \textit{green} and \textit{blue dotted lines} in the right panels for F5 and F6, respectively.} 
\label{fig:power0}
\end{figure*}

\begin{figure*}
\centering
\includegraphics[width = \textwidth]{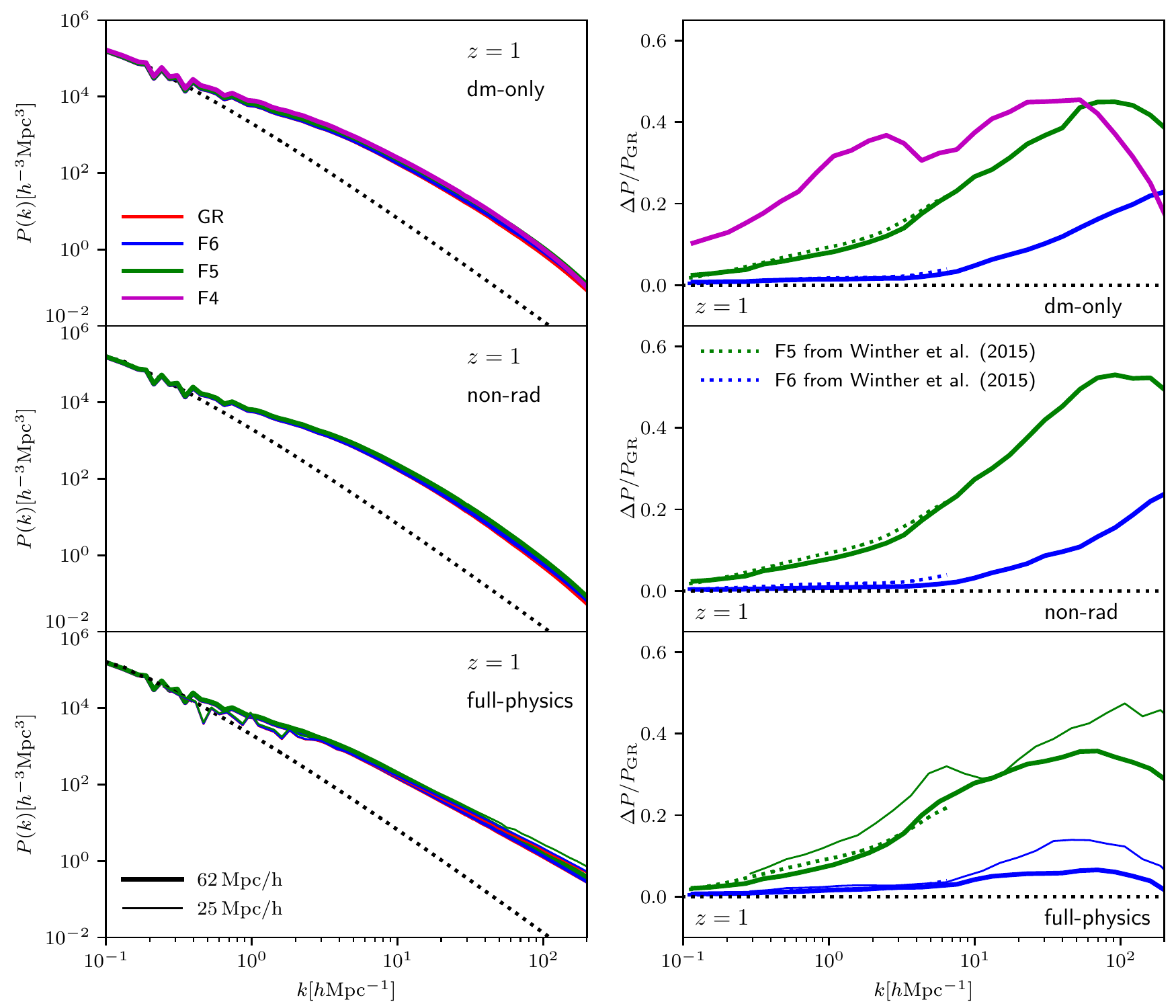}
\caption{Same as Figure \ref{fig:power0} but for $z=1$.} 
\label{fig:power1}
\end{figure*}

\begin{figure*}
\centering
\includegraphics[width = \textwidth]{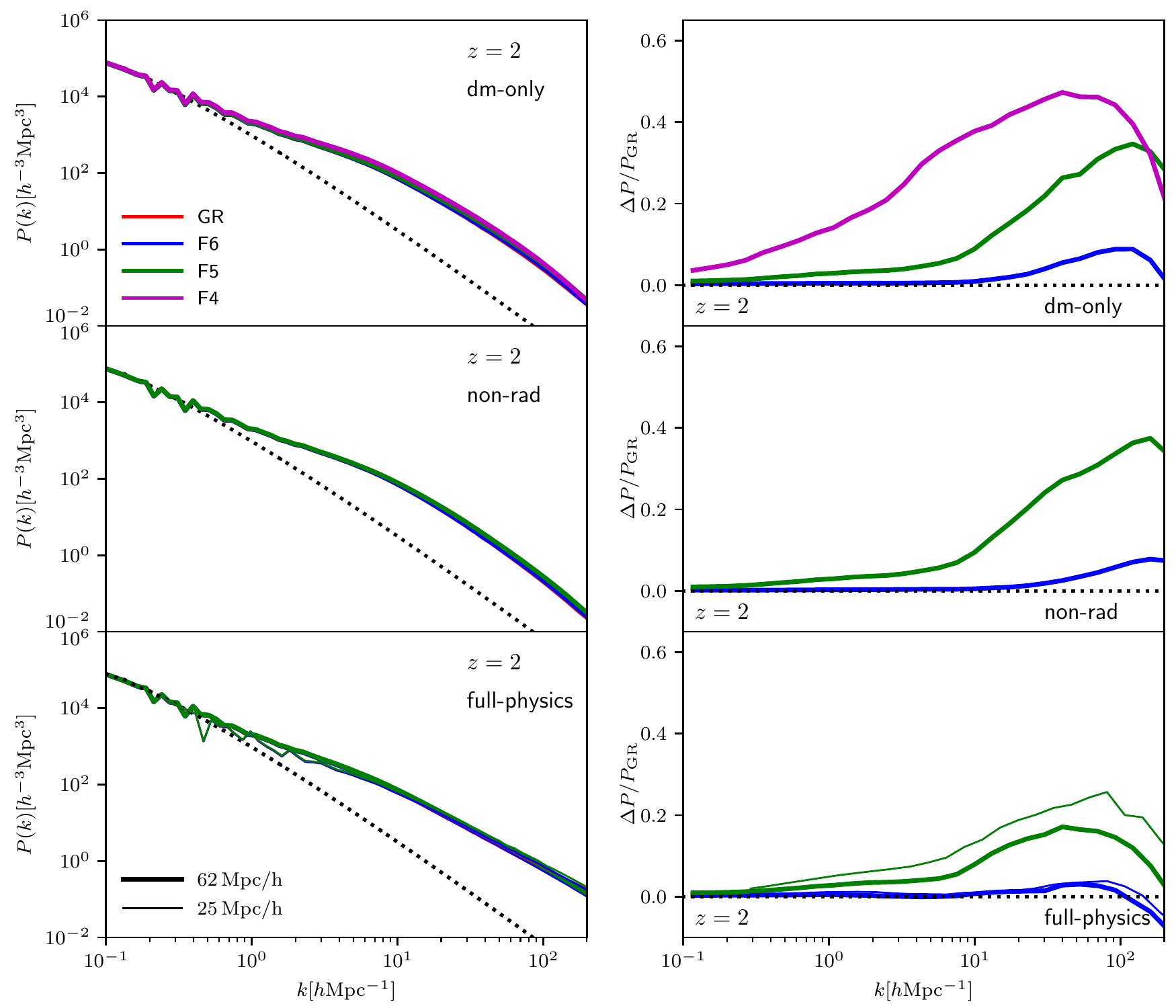}
\caption{Same as Figure \ref{fig:power0} but for $z=2$.} 
\label{fig:power2}
\end{figure*}

\begin{figure*}
\centering
\includegraphics[width = \textwidth]{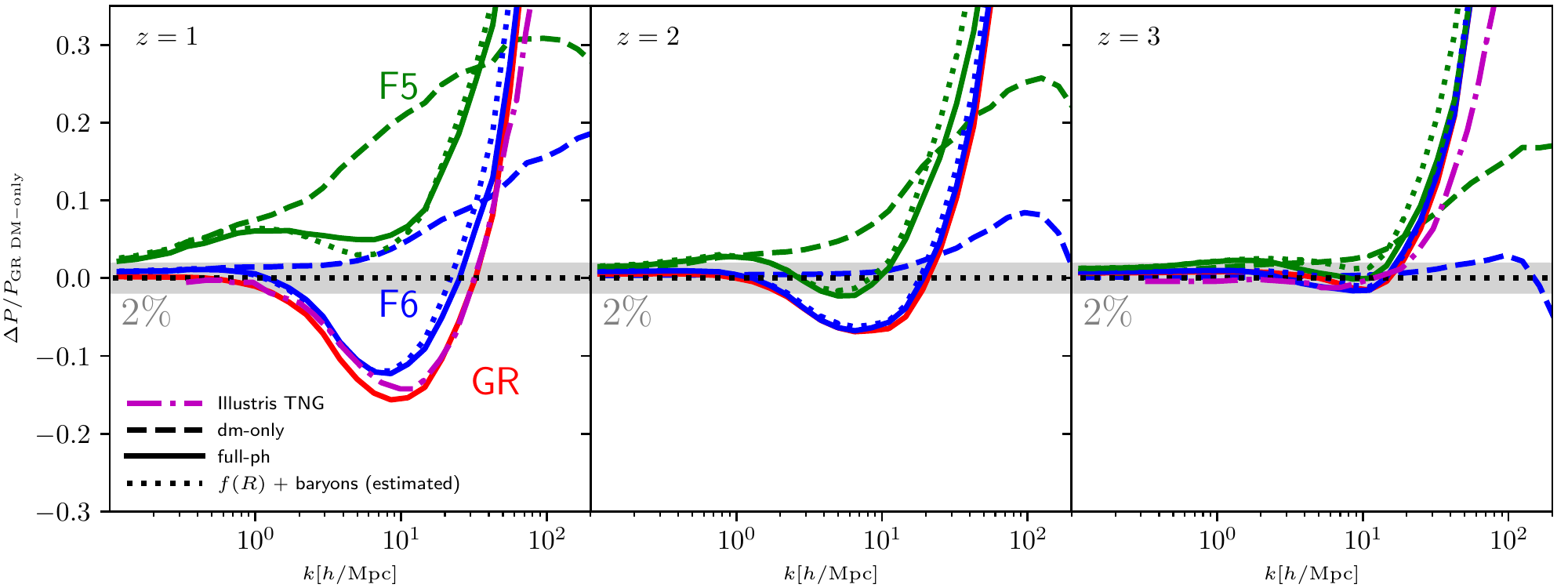}
\caption{The relative difference of different matter power spectra with respect to the power spectrum from our GR DM-only simulation for $z=1$, $z = 2$ and $z=3$ (\textit{left, center, right}, respectively) as shown for z=0 in \protect\cite{arnold2019}. \textit{Red lines} show results for GR, \textit{blue lines} for F6 and \textit{green lines} for F5. \textit{Dashed lines} show results from \fr DM-only simulations, \textit{solid lines} from full-physics simulations. The \textit{dotted lines} show an estimate for the combined $f(R)$ and baryonic effect on the matter power spectrum obtained by adding the relative difference of our GR-full-physics simulation to the relative difference of the \fr DM-only simulation. The dotted horizontal line indicates equality, the grey shaded region shows a $2\%$ margin. The \textit{magenta dash-dotted line} shows the IllustrisTNG GR full-physics result from \protect\cite{springel2018}.} 
\label{fig:power_rd}
\end{figure*}

\begin{figure*}
\centering
\includegraphics[height = 10.2cm]{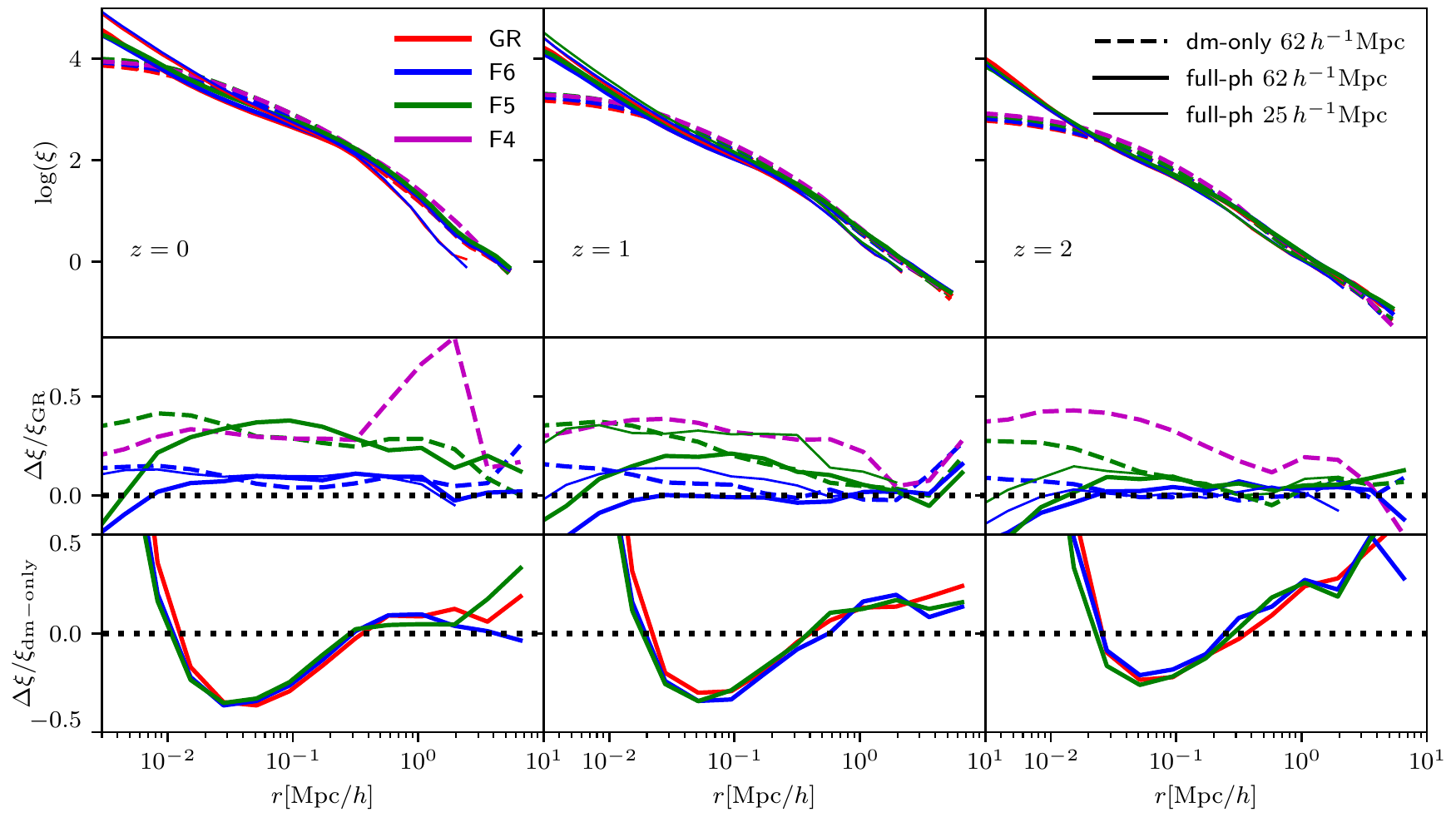}
\caption{The matter correlation function for all simulated cosmologies and hydrodynamical models at redshifts $z=0, 1$ and $2$ (\textit{top panels}). The \textit{middle row} show the relative differences between the \fr results and the \lcdm reference simulation of the corresponding hydrodynamical model. The \textit{lower panels} display the relative difference between the two hydrodynamical simulations and the corresponding DM-only run for each gravity model. The \textit{black dotted vertical lines} indicate zero relative difference. 
In all panels, GR, F6, F5 and F4 results are shown in \textit{red}, \textit{blue}, \textit{green} and \textit{magenta}, respectively. Results from the $62\mpcoh$ dm-only simulations are indicated by \textit{thick dashed lines}, from the $62\mpcoh$ full physics simulations by \textit{thick solid lines} and from the $25\mpcoh$ full physics simulations by \textit{thin solid lines}. 
}
\label{fig:correl_dm}
\end{figure*}

\begin{figure*}
\centering
\includegraphics[height = 10.2cm]{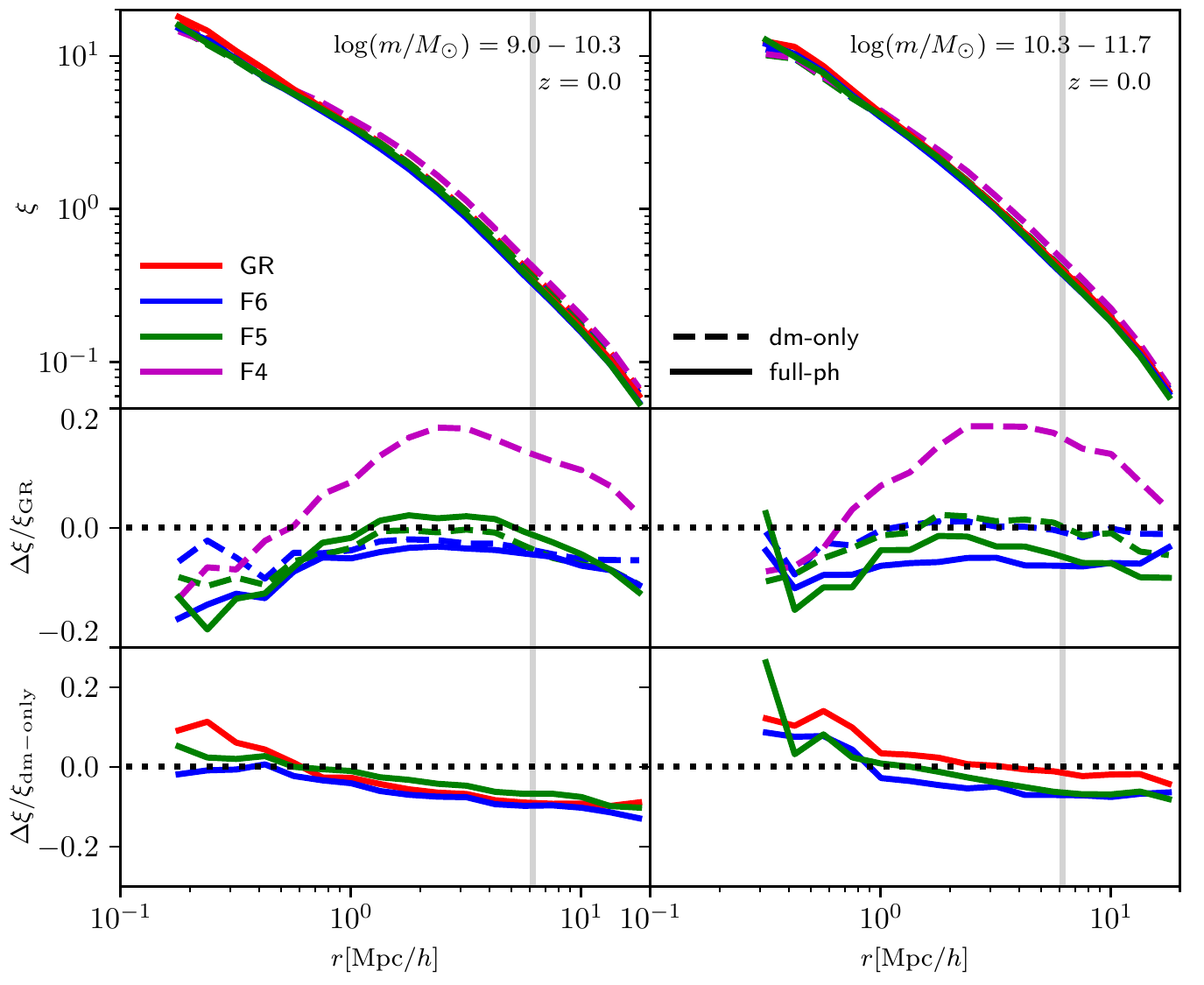}
\caption{The halo-halo correlation function at $z=0$ for two different mass bins (\textit{top panels}) and all simulated gravity and hydro models in the $62\mpcoh$ simulation boxes. GR results are again shown in \textit{red}, F6 and F5 in \textit{blue} and \textit{green}, respectively. \textit{Dashed lines} indicate results from the dm-only simulations, \textit{dash-dotted lines } from the non-radiative hydrodynamical simulations and \textit{solid lines} from the full physics simulations. The \textit{middle row} shows the corresponding relative differences between the \fr simulations and the standard gravity runs. The \textit{lower panels} display relative differences between the hydrodynamical and the collision-less simulations. \textit{Horizontal dotted lines} indicate equality. The \textit{vertical gray lines} indicate the radius beyond which the results are likely affected by the limited box-size of the simulations. }
\label{fig:correl_groups0}
\end{figure*}


\section{Simulations and Methods}
\label{sec:methods}

In this work, we present a detailed analysis of the halo and matter clustering for the SHYBONE (Simulation HYdrodynamics BeyONd Einstein) simulation suite \citep{arnold2019}. 
The simulations were carried out with the \textsc{arepo} hydrodynamical simulation code \citep{springel2010}  and its new modified gravity module \citep{arnold2019}. The module allows to solve the equations for the scalar field (\ref{fRequn}) and the modified Poisson equation for Hu \& Sawicki $f(R)$ gravity to full non-linearity in the quasi-static limit. This way, it can compute the fifth force and capture the effects of the chameleon screening mechanism. The modified gravity solver is based on the $f(R)$-solver in \textsc{mg-gadget} \citep{puchwein2013} but employs the optimised $u^3$ method of \cite{bose2017} and a local time-stepping scheme \citep{arnold2016} for higher efficiency. 

The modified gravity module is combined with the IllustrisTNG galaxy formation model \citep{pillepich2018b, springel2018, genel2018, marinacci2018, nelson2018} implemented in \textsc{arepo}. The TNG model is based on the original Illustris model \citep{vogelsberger2014} and incorporates sub-grid descriptions for a number of astrophysical processes necessary to reproduce a realistic galaxy population in cosmological simulations. These include a description for the growth of super-massive black holes and  feedback from Active Galactic Nuclei (AGN) \citep{weinberger2017}, an algorithm for star formation and stellar feedback, galactic winds as well as a model for the chemical enrichment, UV-heating and cooling of gas. The model uses the magneto-hydrodynamics solver implemented in the \textsc{arepo} code and is tuned to reproduce the observed galaxy stellar mass function, galaxy gas fraction, black hole masses, the cosmic star formation rate density, galaxy sizes and the galaxy stellar mass fraction \citep{pillepich2018} in standard \lcdm simulations.
We did not retune the model for the SHIBONE simulations but confirmed in \cite{arnold2019} that the $f(R)$-effect on these observables is smaller than the uncertainties on the currently available observational data.
 
 The SHIBONE simulations consist of in total 13 simulations carried out for different cosmologies, hydrodynamical models and at two different resolutions. 
 A set of 10 large box simulations uses identical initial conditions with $512^3$ resolution elements for DM and roughly the same number of gas cells (where applicable) in $62\mpcoh$ simulation boxes.
 The large box runs include four DM-only simulations for GR, F6, F5 and F4. Another three $62\mpcoh$ box simulations with a basic non-radiative hydrodynamical model were carried out for GR, F6 and F5. Finally, there are three full-physics simulations using the TNG-model for GR, F6 and F5 using the large box. 
 In addition, there are three full-physics high resolution simulations using the same number of resolution elements in a $25\mpcoh$ box for GR, F6 and F5 (the F5 simulation is run until $z = 1$ only). 
 
 The total computational cost of the simulation suite was about $3$ million core-hours. Depending on resolution, the F6 full-physics simulations are a factor of $3-5$ more expensive than the GR counterpart. For F5 the computational cost of the full physics simulations is about $6-10$ times higher than for GR. The computational overhead in the \fr simulations is partially caused the multigrid-solver which takes $30-80\%$ of the total runtime and partially due to additional global timesteps which are necessary to accurately account for the fifth force effects in the simulations \citep[see][for details on the timestep scheme]{arnold2019}.
 
 All simulations use Planck 2015 cosmology \citep{planck2016} with $n_s = 0.9667$, $h = 0.6774$, $\Omega_\Lambda = 0.6911$, $\Omega_{\rm B} = 0.0486$, $\Omega_{\rm m} = 0.3089$ and $\sigma_8 = 0.8159$. An overview over the simulation suite is given in Table \ref{tab:sims}.

\section{Results}
\label{sec:results}
The results presented in this paper focus on the matter, halo and galaxy clustering and supplement some of the results presented in \cite{arnold2019}. Throughout this work, we use $m_{200}^{\rm{crit}}$ as a mass measure which includes all mass of an object enclosed by a sphere of an average density $\rho = 200\times \rho_{\rm{crit}}$ around the potential minimum of the object identified by the \textsc{subfind} halo finder \citep{springel2001}.

\subsection{The matter power spectrum}
In Figures \ref{fig:power0} - \ref{fig:power2} we present the matter power spectra for all four cosmological models (GR, F6, F5 and F4) studied and the three different hydro models (DM-only, non-radiative hydrodynamics and the Illustris TNG full-physics model) at redshift $z=0, 1$ and $2$. The F4 model was solely simulated for DM-only. We show the absolute values of the power spectra in the left panels while the relative differences between the models are shown on the right hand side (some of the full-physics results have already been shown in \citealt{arnold2019}, we include them here for completeness). The power spectra are corrected for the lack of low-$k$ modes in the initial conditions and also employ a shot-noise correction on small scales. 

The plot shows that the power spectrum is enhanced for the \fr models considered in the DM-only simulations. The relative differences increase towards smaller scales and reach a maximum of roughly $20\%$ at $k \approx 100\hompc$ for the F6 model and of about $50\%$ for the F5 model at the same scale. For the F4 model, the relative difference reaches a maximum of $\approx 60\%$ at $k\approx 1$. The results for F5 and F6 agree very well with those presented in the modified gravity code comparison project \citep{winther2015} for all available redshifts.

The relative difference induced by the considered modified gravity model in the power spectra of the non-radiative simulations is very similar. This is not surprising as these simulations do not include any feedback processes which could be affected by the modifications to gravity and thus lead to back-reactions. The small differences between DM-only and non-radiative simulations appearing at very small scales are caused by the self-interaction of the gas and the resolution difference between the runs (due to the additional gas particles the effective mass resolution is roughly a factor of two better in the non-radiative runs). 

In the lower panels we show the power spectra for the full-physics simulations which has already been partly presented in \cite{arnold2019}. The relative difference in the power spectrum is affected by the complicated interplay between AGN and supernova feedback and \fr. This causes the relative difference to be larger at intermediate scales $k\approx10\hompc$. 

The results from the large and the small simulation boxes in the left panels agree on intermediate scales. As one would expect, the small simulations lack modes on large scales due to their limited boxsize while the large boxes are affected by resolution effects on small scales in the plot. 
Comparing the relative difference between \fr and standard gravity from the two simulation boxes we find the different resolution simulations to agree within a few percent for scales larger than $k=10\hompc$. A moderate resolution dependence is expected as the original TNG simulations show differences in the power spectrum between different resolutions \citep{springel2018}. The discrepancies of up to $5\%$ at $k=1\hompc$ nevertheless show that it will be very hard to find fitting formulae for the power spectrum enhancement in \fr which are accurate to $1\%$ at this or even smaller scale and include baryonic effects. 

In Figure  \ref{fig:power_rd} we show the relative difference of matter power spectra from full-physics and DM-only simulations with respect to a GR DM-only simulation for redshift $z = 1, 2$ and $3$ as already shown for $z=0$ in \cite{arnold2019}. The plot allows to study the degeneracy in the matter power spectrum between baryonic feedback and \fr at different redshifts. As expected from previous works \citep{springel2018}, the power spectrum in full-physics simulations is significantly enhanced on very small scales ($k>30\hompc$) with respect to  GR DM-only. This is true for all gravity models and redshifts. For redshifts $z<3$, the baryonic feedback as implemented in the TNG model causes an additional suppression of the power spectrum at scales around $k = 10\hompc$. The \fr effect on top of this is small at $z=3$ but becomes size-able at lower redshift.

 In \cite{arnold2019} we find that the back reaction effect between \fr and baryonic feedback at $z=0$ is negligible for the F6 model but plays a non negligible role for F5. To study the behaviour of this back-reaction at higher redshift we plot estimates for the combines modified gravity and baryonic effect, which is obtained by adding the GR-full physics result to the individual $f(R)$ DM-only results, as dotted lines in Figure \ref{fig:power_rd}. 
 Comparing these estimates to the full physics results (solid lines) for F6 and F5, it turns out that the same is true for $z=1$. At $z=2$ and $z=3$ the back-reaction effect is negligible for both models. This is easily explained by the background evolution of the scalar field. The F6 model screens the centres of objects which carry an AGN (i.e. massive haloes) very effectively at all redshifts. The AGN feedback process is thus not effected by the changes to gravity, leading to no back-reaction effects. In F5, the centres of many AGN-hosts are only screened at high redshift, suppressing back-reaction effects at $z>1$. At $z=1$, the centres of (sufficiently many) AGN hosting haloes become unscreened in F5. The inflows onto and outflows from the black hole therefore take place in an environment with enhanced gravity, affecting the AGN feedback efficiency which is visible in the matter power spectrum.

 \subsection{Matter correlation functions}
 As a complementary result, we show the (total) matter correlation function for the DM-only (dashed lines) and the full-physics (solid lines) simulations at redshift $z = 2, 1$ and $0$ in Figure \ref{fig:correl_dm}. Thick lines illustrate results from the large simulation boxes, thin lines from the small box runs. 
Along with the absolute values (top panels) we show the relative difference between the \fr results with respect to the \lcdm cosmology simulations (middle panels) and the relative differences between full-physics and DM-only for each of the gravitational models (bottom panels).

As expected, the plots reproduce the clustering behaviour observed for the power spectra above. At large and intermediate scales the relative differences between \fr and standard gravity are only mildly affected by feedback and hydrodynamical processes in the full physics simulations. The correlation functions from the DM-only and full-physics simulations thus show a similar relative difference between modified gravity and GR at these scales. At very small scales ($r < 2\times10^{-2}\mpcoh$), the relative differences for the full-physics simulations obey the same downturn as for the power spectrum which is not present for DM-only. 
The downturn at small scales is already present at high redshift and can again be explained by the interplay between AGN feedback and enhanced forces due to modified gravity:
The enhanced gravitational forces in \fr alter the inflow of matter onto the AGN and the center of massive objects but also affect the outflows from the AGN. 
The overall relative difference is larger at smaller redshifts. This is a result of the time dependence of the background field in \fr which leads to a lower screening threshold at high $z$.
As for the power spectra, the \fr effect on the correlation function is mildly resolution dependent, showing the complexity of the interplay between baryonic processes and modified gravity. 

Interestingly the correlation functions reaction to hydrodynamical processes is relatively independent of the gravitational theory. The relative differences between the full-physics and DM-only simulations shown in the lower panels all follow a similar pattern: at very small scales matter shows enhanced clustering in hydro simulations while the clustering is suppressed on scales of $k = 0.1\hompc$. At larger scales the correlation functions are enhanced again. This pattern is observed at all three redshifts while the large scale enhancement is larger at larger redshift. 
The similar reaction of all three cosmological models to feedback is expected from the results presented in \cite{arnold2019} where we report that back reaction effects  between (primarily) AGN-feedback and \fr are negligible for the F6 model and lead to only few-percent differences for F5. 

\begin{figure*}
\centering
\includegraphics[height = 10.2cm]{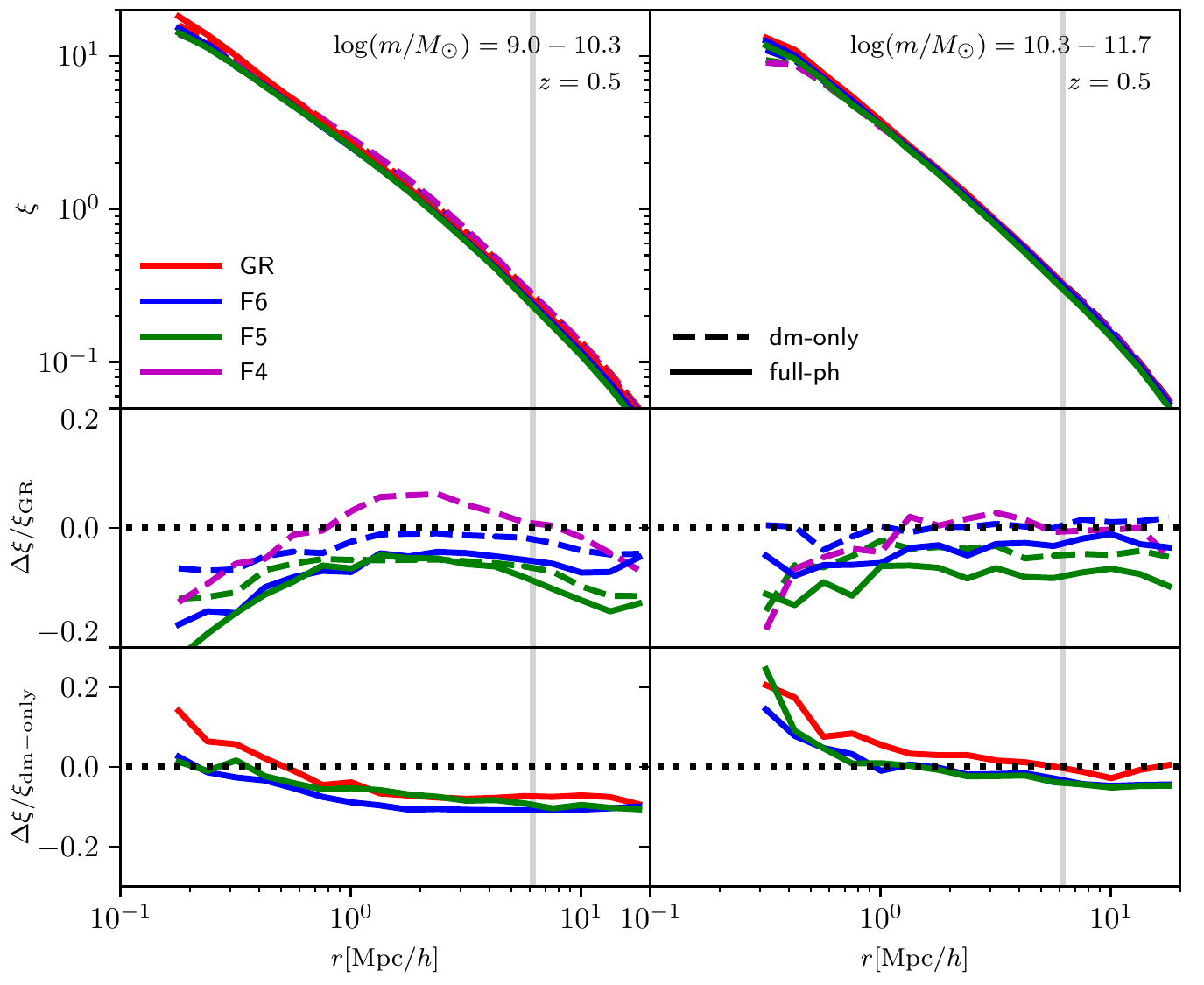}
\caption{Same as Figure \ref{fig:correl_groups0}, but for $z=0.5$.}
\label{fig:correl_groups05}
\end{figure*}

\begin{figure*}
\centering
\includegraphics[width = \textwidth]{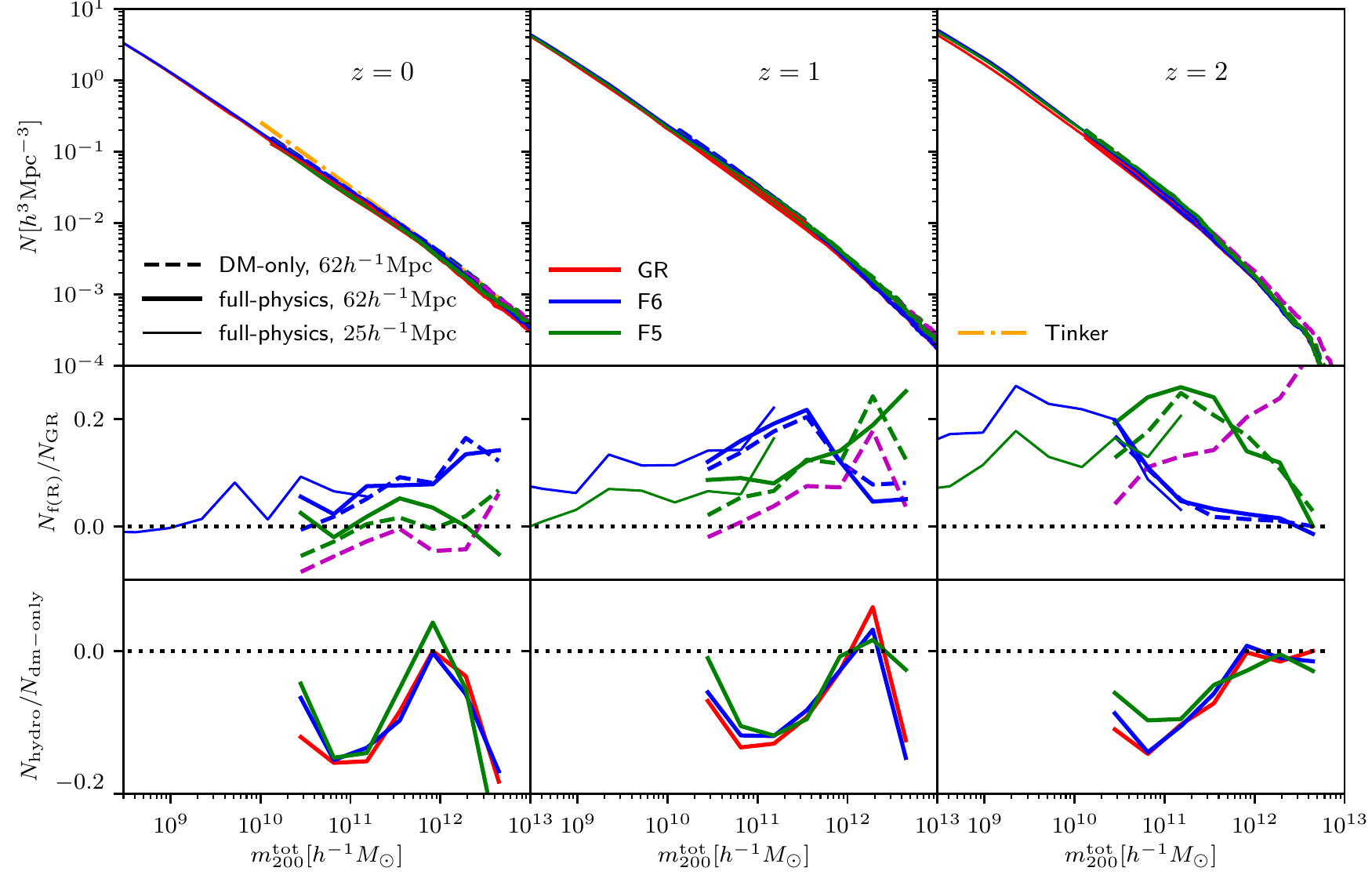}
\caption{The halo mass function for the full-physics and DM-only simulations performed within this project at redshifts $z=0$ (\textit{left panels}), $z=1$ (\textit{center panels}) and $z=3$ (\textit{right panels}). The \textit{top row} shows the absolute value of the mass functions measured in terms of $m_{200}^{\mathrm{crit}}$. Results for a \lcdm cosmology are displayed in \textit{red}, the F6 and F5 models are shown as \textit{blue} and \textit{green lines}, respectively. \textit{Dashed lines} represent the mass function in the DM-only simulations,  \textit{thick solid lines} in the $62\mpcoh$ full physics simulations and \textit{thin solid lines} in the $25\mpcoh$ full physics simulations (for F5, the small box results are only shown for $z>0$). 
\textit{Dash-dotted orange} lines show predictions from \protect\cite{tinker2010}.
The panels in the \textit{middle row} shows the relative difference between the \fr mass function and the corresponding GR result. The \textit{bottom row} displays the relative difference between the mass functions from the hydrodynamical simulations with respect to the dm-only counterpart. The \textit{horizontal black dotted lines} indicate equality. }
\label{fig:massfunc}
\end{figure*}

\begin{figure*}
\centering
\includegraphics[width = \textwidth]{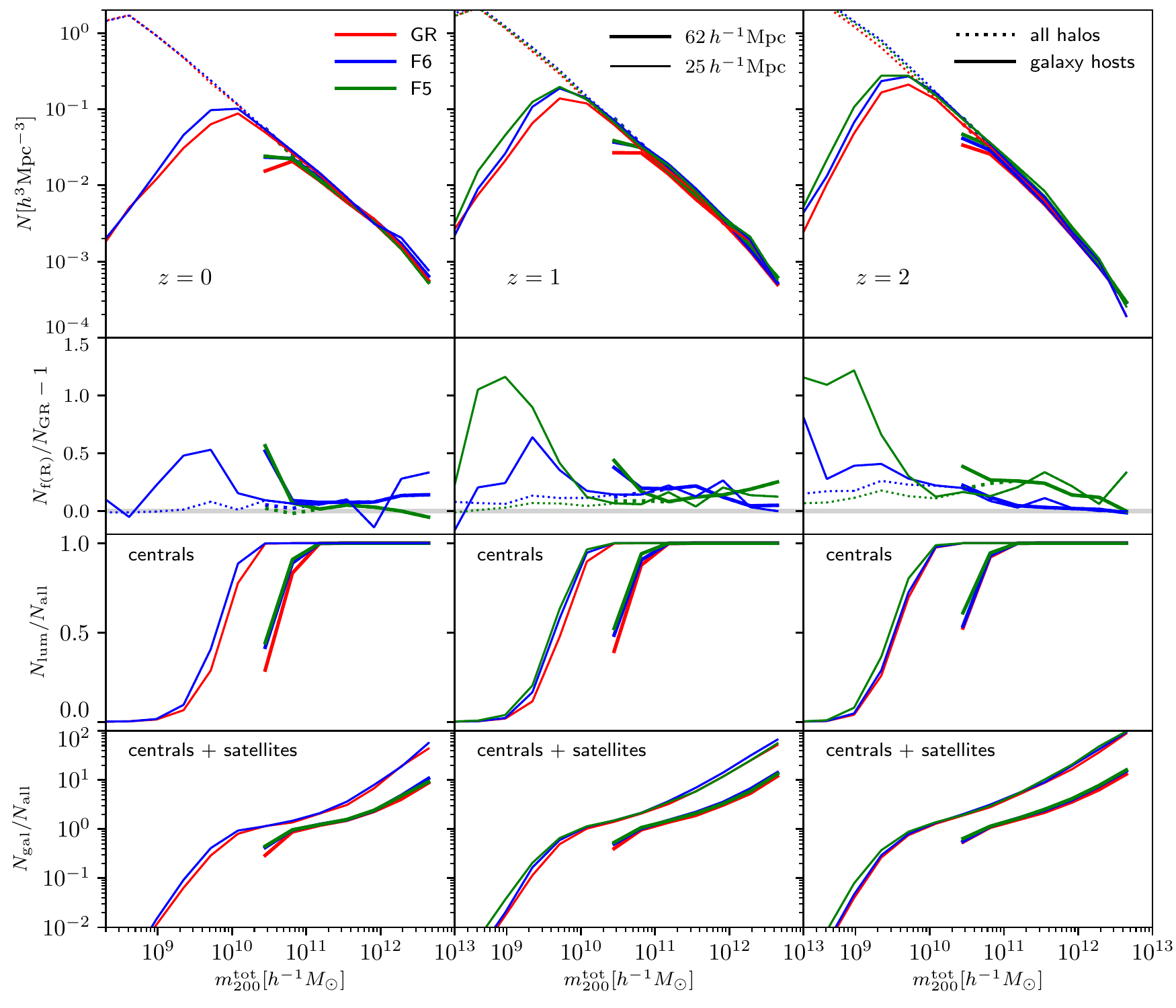}
\caption{The galaxy host halo mass function (\textit{top panels, solid lines}) plotted against total halo mass within $r_{200}^{\rm{crit}}$ at redshifts $z=0$ (\textit{left panels}), $z=1$ (\textit{center panels}) and $z=2$ (\textit{right panels}) compared to the halo mass function (\textit{dotted lines}). Results for the GR simulations are shown as \textit{red lines}, F6 and F5 results as \textit{blue} and \textit{green} lines, respectively. \textit{Thick lines} display results from the $62\mpcoh$ boxes, \textit{thin lines} results from the $25\mpcoh$ boxes (small box F5 only for $z = 1, 2$). The \textit{upper middle row} display the relative differences between the \fr host halo mass functions and the GR result. The \textit{lower middle row} show the fraction of luminous centrals to haloes. For this plot, we consider all objects whose stellar mass is at least $10^{-3}$ times their total mass. We nevertheless found, that the actual mass cut does not have a significant impact on the relative difference between the modified gravity and the \lcdm mass functions. The \textit{bottom row} shows the number of galaxies (including centrals and satellites) per group.}
\label{fig:host_halo_mf}
\end{figure*}

\begin{figure}
\centering
\includegraphics[width = \columnwidth]{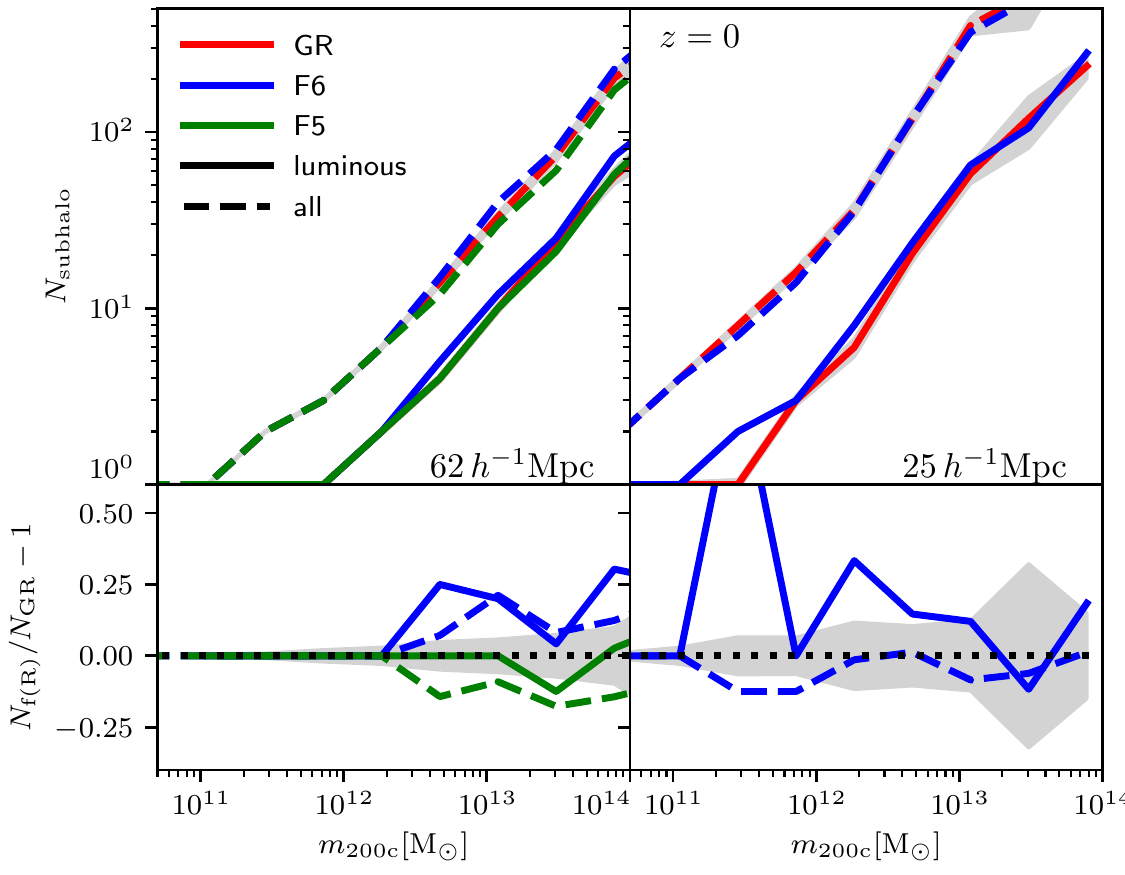}
\caption{\textit{Top panel:} The number of sub-haloes as a function of host halo mass at $z=0$. The \textit{dashed lines} show results considering all bound substructures within a group found by the \textsc{subfind} algorithm, the \textit{solid lines} consider only luminous substructures, i.e. substructures with a stellar mass $m_{\rm stars} > 0$. The \textit{blue lines} show results for a \lcdm cosmology, \textit{red} and \textit{green lines} for the considered F6 and F5 cosmologies. The grey shaded regions indicate the estimated error for the standard gravity results.
The \textit{lower panel} displays the relative difference between the GR and modified gravity results. The \textit{black dotted line} indicates zero difference.
} 
\label{fig:luminous_subs}
\end{figure}

\begin{figure*}
\centering
\includegraphics[width = \textwidth]{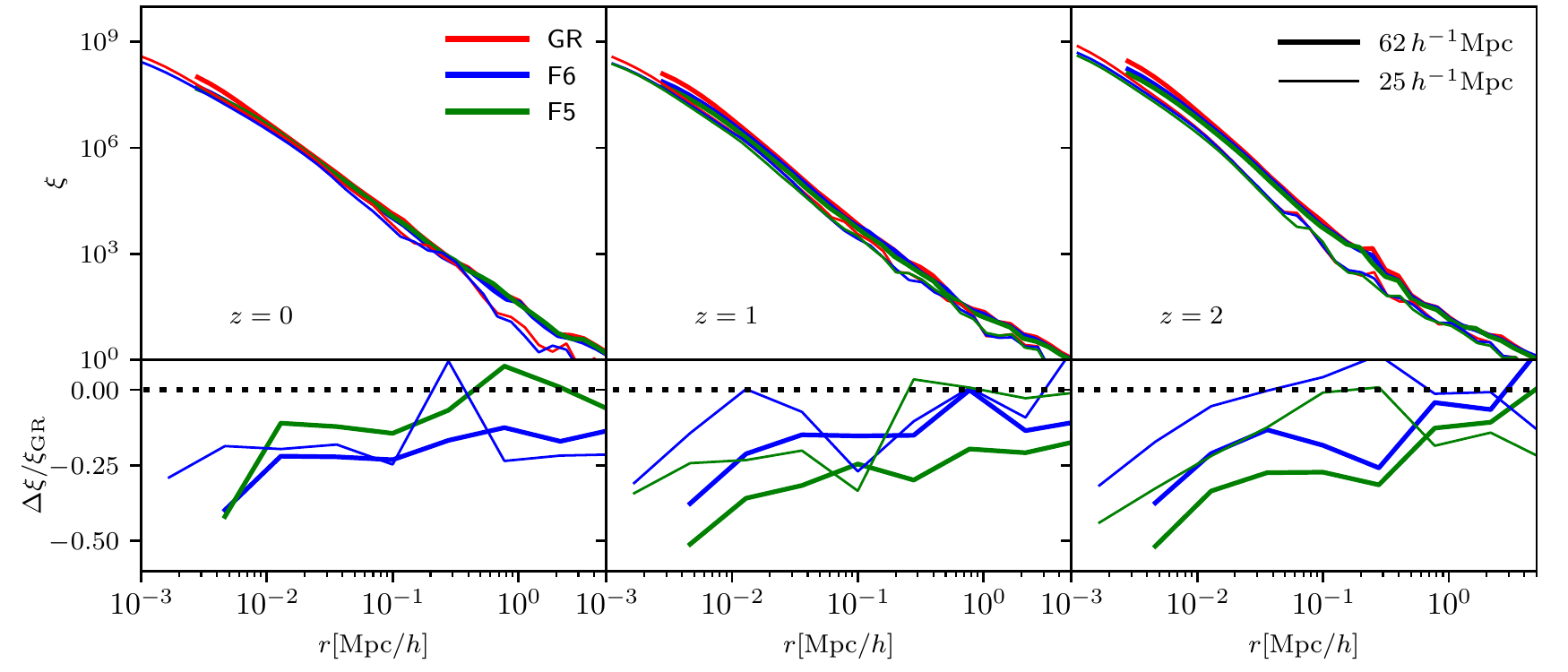}
\caption{The twopoint correlation function of the stars (\textit{top panels}) for standard gravity (\textit{red}), the F6 model (\textit{blue}) and F5 (\textit{green}). \textit{Thick lines} show results for the large simulation boxes, \textit{thin lines} for the small boxes. The \textit{lower panels} display relative differences between the two \fr simulation results and the fiducial \lcdm correlation function. Results for $z=0$ are shown in the \textit{left panels}, for $z=1$ in the center panels and for $z=2$ in the right panels. The \textit{horizontal dotted lines} in the \textit{lower panels} indicate equality.}
\label{fig:correl_stars}
\end{figure*}

\begin{figure*}
\centering
\includegraphics[width = \textwidth]{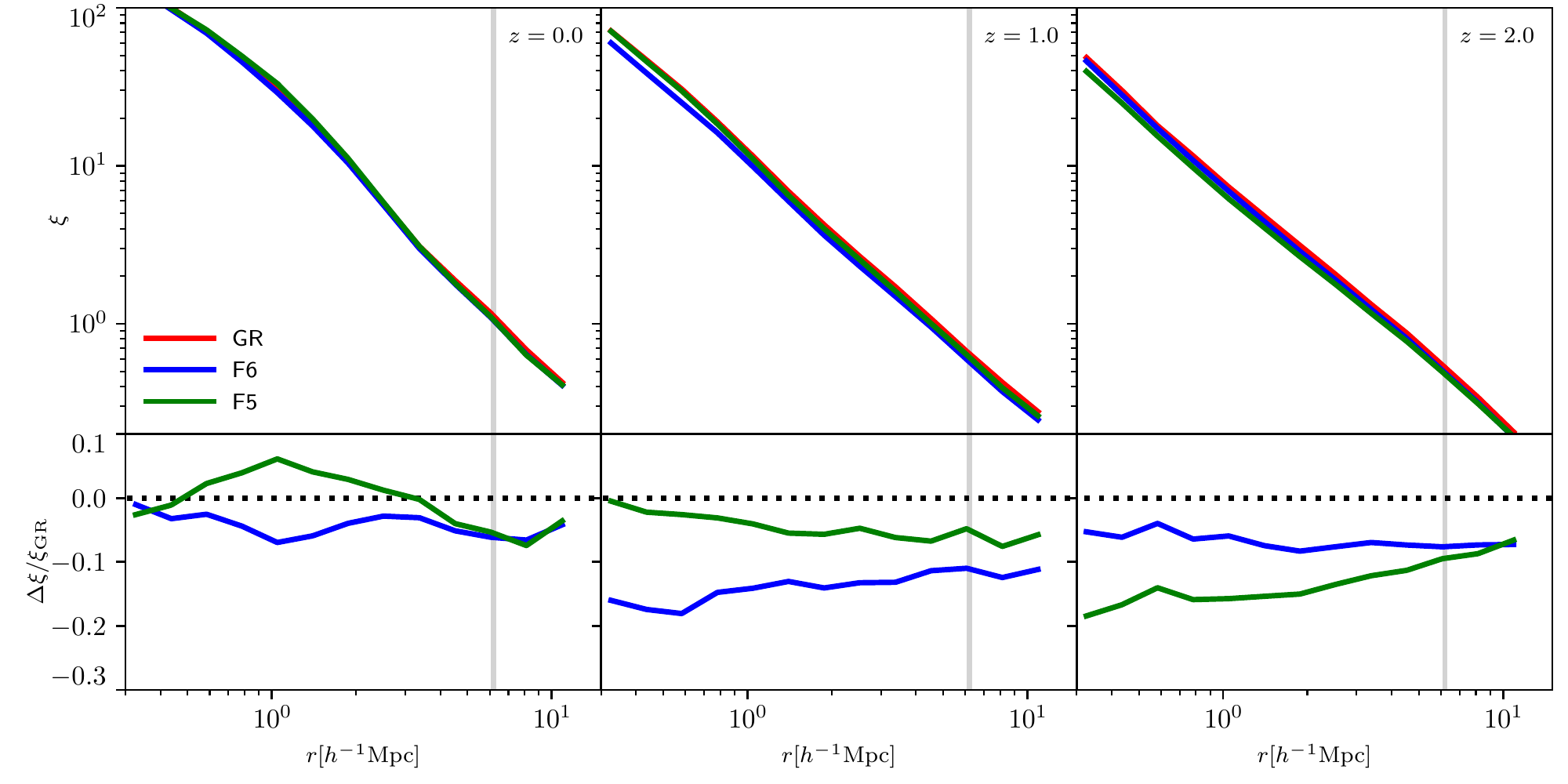}
\caption{The galaxy auto-correlation function in the $62\mpcoh$ simulations at redshift $z=0$ (\textit{left}), $z=1$ (\textit{center}) and $z=2$ (\textit{right}). \textit{Red lines} show results for GR, \textit{blue lines} for F6 and \textit{green lines} for F5 in the \textit{upper panels}. The lower panels show relative differences between the two \fr  and standard gravity simulations. The \textit{horizontal dotted lines} in the lower panels indicate equality. The \textit{vertical gray lines} indicate the radius at which the results start to be affected by the limited boxsize of the simulations (i.e. $1/10$ of the boxsize.}
\label{fig:correl_galaxies}
\end{figure*}

\subsection{Halo auto-correlation}
In Figure \ref{fig:correl_groups0} we analyse the halo-halo auto-correlation function for (central) haloes of two different mass bins at redshift $z=0$. In the upper panels the figure displays the absolute values of the correlation function for the two \fr models studied and for the \lcdm reference simulation for both the full-physics and the DM-only runs. The middle row shows the relative differences between the correlation functions in the $f(R)$ simulations relative to the corresponding GR runs while the lower panels show the relative difference between the full-physics simulations and their DM-only counterparts for all three cosmological models. Because of their small volume, the $25\mpcoh$ simulations lack statistical power, we therefore do not show these results for the halo correlation function. The radial range for which the correlation functions might be affected by boxsize (i.e. scales larger than $1/10$ of the boxsize) for the large simulation boxes is marked by the vertical gray lines in the plots. 

The middle row in Figure \ref{fig:correl_groups0} shows that the halo correlation function is mildly suppressed in the F6 and F5 model simulation relative to GR on scales larger than $1\mpcoh$. On smaller scales, the suppression is stronger. These results for F5 are consistent with those presented in \cite{arnold2018}. 
In the F4 model, the halo correlation function shows a different behaviour. The clustering of haloes is enhanced by about $15\%$ on intermediate scales for both mass bins and drops to $\approx 0$ on for $r < 1 \mpcoh$. 

The panels in the lower row show that the auto-correlation function of the haloes is only mildly affected by baryonic effects and feedback for GR, F6 and F5. The the halo correlations in three gravity models react in a similar way to the baryonic processes, showing a few percent suppression of clustering on larger scales ($\approx5\mpcoh$) and a small enhancement on small scales. 

In Figure \ref{fig:correl_groups05} we present a similar analysis, but for $z=0.5$. 
The differences between the F5 and F6 models and GR are larger at this redshift, showing a $10\%$ suppression of halo clustering for the F5 full Physics simulations relative to GR. For the F4 model, the difference to GR is much smaller. 

The unscreened objects experience the by a factor of $4/3$ enhanced gravitational forces in \fr and therefore grow faster compared to standard gravity. Therefore, haloes which correspond to less clustered lower initial density contrast peaks in the density field will now contribute to the higher mass bins in the correlation function and thus lead to a suppression of clustering. This effect is strongest for halo masses which have recently become unscreened, leading to a stronger suppression of clustering in the considered mass bins for the F6 model compared to F5 at $z=0$.
For the F4 model haloes have been unscreened for a long time at both redshift $z=0$ and $z=0.5$ already. The suppression due to faster growth is therefore compensated by the larger merger and satellite capture rates leading to an enhanced clustering of haloes. At $z=0$, this effect leads to an enhancement in clustering for the F4 model compared to GR.

\subsection{The halo and galaxy-host-halo mass function}
The halo mass function is shown in Figure \ref{fig:massfunc}. For this plot we include all bound objects identified by \textsc{subfind} in the total mass density field. The masses are given in terms of $m_{200}^{\rm{crit}}$. In the top panels we compare absolute value of the mass functions measured from our DM-only and full-physics simulations to the theoretical fitting formulae of \cite{tinker2010} for $z=0$ and $z=1$. As expected, the Tinker fitting formula is in very good agreement with  the results from our DM-only simulations. 

In the middle panels we show the relative difference between the \fr models and the results from the \lcdm simulation. The relative differences in the DM-only simulations follow the trend expected from previous works \citep{schmidt2010, winther2015, shi2015, arnold2018}: The halo mass function in \fr is enhanced with respect to GR. The enhancement can be described by a triangular shaped function. Above a certain mass, the chameleon screening mechanism is active and the objects are thus unaffected and the mass function is the same as in a standard gravity simulation. The triangular enhancement can be observed in the mass range where objects have just become unscreened. Due to enhanced gravitational forces, mass accretion happens faster and thus shifts objects in the halo mass function towards higher masses leading to an excess of objects below the screening threshold. This threshold depends on the current background field and redshift. The location of the peak of the enhancement in terms of mass thus depends on the model parameter and redshift. As one can see from the plot, our simulation results match exactly this behaviour. At $z=2$, the objects at the high mass end ($m_{200}^{\mathrm{cirt}} > 10^{12.5}h^{-1}M_\odot$) are not yet affected by the changes to gravity while there is a significant ($\approx 20\%$) larger number of haloes present at this mass at $z=1$ for the F5 model. It is also apparent that the F5 model affects larger mass haloes compared to the F6 model at a given redshift due to the different screening thresholds.

The relative differences between the considered modified gravity models and GR in the full-physics simulations are similar to the DM-only runs. The results from  the large and small simulation boxes broadly agree in the overlap region. Small differences appear due to the limited number of higher mass objects in the $25\mpcoh$ simulation box.

The lower panels of Figure \ref{fig:massfunc} show the relative difference between the full-physics TNG model simulations and the DM only runs for the three considered cosmologies in the large box simulations. The results show that there is an about $15\%$ lower number of intermediate mass haloes in the mass range $10^{11}-10^{12}\,h^{-1}M_\odot$ in the full-physics simulations relative to DM-only. This is consistent with \citep{springel2018} who show that haloes of this mass range are between $5\%$ and $10\%$ less massive compared to their DM-only counterparts. 
The suppression is strongest where feedback is most efficient, i.e. for masses $m_{200}^{\mathrm{crit}} < 10^{12}h^{-1} M_\odot$ for stellar feedback and for $m_{200}^{\mathrm{crit}} < 10^{12}h^{-1} M_\odot$ for AGN feedback (for $z<2$). At the peak of the stellar mass fraction ($m_{200}^{\mathrm{crit}} \approx 10^{12}h^{-1} M_\odot$, see \citealt{arnold2019}) the combination of both effects is least efficient and the mass function in the full-physics simulations is similar to the DM-only counterparts. 
The effects are similar across the three models. The only differences occur at the low mass end of the plot where the haloes in the GR simulations are affected more strongly by the hydrodynamical effects than in the \fr runs. This effect might nevertheless be caused by the limited resolution of the $m_{200}^{\mathrm{crit}} \approx 10^{10}\,h^{-1}M_\odot$ haloes in our simulations.

Our results on the halo mass function show, that the \fr and feedback effects can be treated independently for halo masses $m_{200}^{\mathrm{crit}} > 5\times10^{10}\,h^{-1}M_\odot$. There is no sign of a strong back reaction between modified gravity and baryons for the halo mass function.

 In order to better understand the differences in galaxy formation between standard gravity and the considered modified gravity models, we plot the galaxy host halo mass function in Figure \ref{fig:host_halo_mf}. For the plot, all gravitationally bound (central) objects identified by \textsc{subfind} which have a stellar mass larger than $10^{-3}$ times the total mass of the object (both measured within $r_{200}^{\rm{crit}}$) are binned according to their mass. In the top panels we show the absolute values of the host halo mass function for $z = 0, 1$ and $2$.  The Figure also shows the halo mass function, for comparison.
 As most of the high mass haloes ($m_{200}^{\mathrm{cirt}} > 10^{11}h^{-1}M_\odot$) host a galaxy, the galaxy host halo mass function follows the trend of the halo mass function at the high mass end of the plot. At lower masses not all objects host a galaxy, which leads to a suppression of the host halo mass function relative to the halo mass function. For haloes with a mass lower than $10^{10}\,h^{-1}M_\odot$, the probability of hosting a halo is very low \citep{pillepich2018b}, there are thus only few galaxies left at this mass range.  
 
The second row from the top shows the relative differences between the \fr simulations and the \lcdm results.
Given that all high mass haloes host at least one galaxy it is not surprising that the relative differences between the gravity models in the host haloes mass function follow those of the halo mass function for $m>5\times 10^{10} h^{-1} M_\odot$ and are small.
Interestingly, relatively large differences between the models occur at the low mass end of the plot, where the host halo mass function starts to differ from the halo mass function. 
For the F5 model, about $100\%$ more haloes with masses around $10^{9.5} \,h^{-1}M_\odot$ are populated with galaxies relative to \lcdm at redshift $z=1$ and $z=2$. 
In F6, the relative difference is $\approx 50\%$ at $z=0$ and $1$ and about $25\%$ at redshift $z=2$.

 One of the main reasons for the larger galaxy number in \fr is the effect of the increased gravitational forces on the gas density within haloes: Low mass objects become unscreened already at high redshifts. The gravitational forces within the objects are thus a factor of $4/3$ higher compared to GR. This leads to higher gas densities and more efficient gas cooling within the objects. The overall denser and colder gas is finally more likely to form stars, leading to an earlier onset of star formation within low-mass objects and a larger fraction of haloes hosting galaxies in \fr compared to standard gravity. 
The same effect is also visible in the power spectrum of the stars \citep[see][]{arnold2019} and the stellar correlation function which we will discuss below. 

The fraction of haloes populated by central galaxies is displayed in the third row of Figure \ref{fig:host_halo_mf}. It shows a resolution dependence in the transition region, where the fraction of populated galaxies drops from $1$ to $0$. In the large simulation box, the star formation rates are lower due to the lower mass resolution (\citealt{pillepich2018}, \citealt{arnold2019}). Objects with halo masses between $10^{10} \,h^{-1}M_\odot$ and $10^{11} \,h^{-1}M_\odot$ are consequently less likely to form a galaxy in the large box. The transition from no haloes populated with galaxies to all haloes populated with galaxies thus takes place at higher mass.

The total number of galaxies per central halo is displayed in the lower row of Figure \ref{fig:host_halo_mf}. Again, all large mass haloes ($m_{200}^{\mathrm{crit}} > 10^{10}\, h^{-1}M_\odot$) host at least one galaxy, but with increasing mass more and more subhaloes get populated with galaxies as well. The \fr simulations show a small increase in subhalo number relative to GR for the same reasons as explained above. 

\subsection{Subhalo statistics}
In Figure \ref{fig:luminous_subs} we plot the number of subhaloes per halo as well as the number of (satellite) galaxies (or luminous subhaloes) identified by \textsc{subfind} as a function of halo total mass within $r_{200}^{\rm{crit}}$ at $z=0$. For both estimates, we also show the standard deviation of the mean within the bins for the GR result as the gray shaded region. The left panels show results from the large box, the right panels from the small box. 
Relative differences between the \fr models and GR are displayed in the lower panels. 
The figure shows that the relative differences between both \fr models and \lcdm are relatively small and broadly within the estimated errors at all redshifts. Given the noisy results we therefore conclude that our simulation boxes are too small in volume to provide enough statistics for a reliable study of the at most small effect of \fr on the subhalo number. The spike in the relative difference between F6 and GR for the small simulation box is likely to be caused by small fluctuations in the overall very low median number of luminous subhaloes (2 galaxies in F6 compared to 1 galaxy in GR) and possible resolution effects at these halo masses.

\subsection{Stellar distribution}
In order to better understand the large changes in the distribution of stars and neutral, cold gas at high redshift presented in \cite{arnold2019} we plot the two-point correlation function of the stars in Figure \ref{fig:correl_stars}. 
The top panels show the absolute values of the correlation functions at redshifts $z=0$ (left), $z=1$ (center) and $z=2$ (right). The lower panels show the relative difference between \fr and the \lcdm cosmology.  
As expected from \cite{arnold2019}, the stars are less correlated in both $f(R)$-models compared to GR.  Even for the F6 model, which is a relatively weak modification of gravity and well consistent with most cosmological tests, the relative differences reach $25\%$ on small scales already at $z=2$. This is particularly interesting as the small background scalar field at $z=2$ allows very efficient chameleon screening. The effect can nevertheless be explained through the interplay of enhanced gas density and the population of low-mass objects with stars. 
At high redshifts, low mass objects become unscreened first. Within these haloes, the gravitational forces are increased due to \fr. This leads to higher gas densities in the objects which allow for more effective gas cooling and thus cause a higher star formation rate. Consequently, more low mass haloes will be populated with stars compared to GR at early times in \fr (see Figure \ref{fig:host_halo_mf}). 
These objects correspond to lower density contrast peaks in the initial conditions of the simulation (or in the density field at recombination) and are thus less correlated than their more massive counterparts. Populating these objects with stars, will consequently lead to a suppression of the stellar twopoint correlation function or the stellar power spectrum. As shown in \cite{leo2019} this effect is also important for the neutral hydrogen distribution and is therefore measurable via 21cm intensity mapping. 

The absolute values of the stellar correlation function in the top panels show a resolution dependence. At large radii, the small box cannot reliably reproduce the correlation function due to its limited volume. At small radii, the large box does not provide enough resolution for an accurate measurement. The relative differences between \fr and standard gravity are also affected by resolution, the results from the large simulation box are not fully converged for $z>0$ due to the dependence of the star formation rate on resolution in the TNG model \citep{pillepich2018}. At $z=0$, the results from the large and small simulation box agree for the F6 model. 

\subsection{The galaxy correlation function}
The galaxy correlation function within our $62\mpcoh$ simulations is displayed in the upper panels of Figure \ref{fig:correl_galaxies}. The lower panels show the relative difference between \fr and GR. Due to the relatively small dynamic range of our simulations, we do not split the galaxy correlation function into individual mass or magnitude bins. As one can see from the Figure the galaxy correlation is suppressed by up to $15\%$ in \fr relative to standard gravity at $z=1$ and $z=2$. At $z=0$ the F6 model shows a $\approx 5\%$ suppression while clustering is enhanced for the F5 model. With that, the galaxy clustering roughly follows the \fr effect on the halo clustering in the full physics simulations. As a cautionary remark, we nevertheless mention that these results can only indicate a trend and that simulations with a larger dynamical range would be needed if a direct comparison to observational data is intended. The small volume of our small simulation boxes does furthermore not allow to check the large box results for the galaxy correlation function for convergence. 


\section{Summary and Conclusions}
\label{sec:conclusions}
We present an analysis of the matter, halo and galaxy clustering in \fr using the SHYBONE simulation suite \citep{arnold2019}, a set of high resolution full-physics hydrodynamical simulations in \fr.
The simulations enable us to study  the combined effect and interaction of baryonic feedback and \fr. 
In order to disentangle both in our full-physics hydrodynamical simulations, we compare the results to both DM-only runs using the same initial conditions and fiducial \lcdm cosmology reference runs. We also compare full-physics \fr simulations at different resolution to test our results for convergence.
Our findings can be summarized as follows:

\begin{itemize}

\item The 3D matter power spectrum is enhanced in \fr. The enhancement in our DM-only simualtions meets the expectations from previous works and shows that the power spectrum enhancement in \fr can be predicted at high accuracy in DM-only simulations.
Our full physics simulations show that the interplay of baryonic processes and \fr leads to an additional enhancement of power on intermediate scales while baryons suppress the enhancement at very small scales. Comparing different resolution full-physics simulations we find that the results for the power spectrum are converged at few-percent level. The remaining differences will nevertheless make it very difficult to calibrate analytical models for the power spectrum enhancement which include baryonic effects if $1\%$ accuracy is desired for scales with $k > 1\hompc$.
\item As already found in \citep{arnold2019}, stars are less correlated in our \fr simulations compared to a \lcdm cosmology on small scales.  Our results on the galaxy host halo mass function suggest that this effect is caused by higher gas densities in low mass objects in modified gravity which cause enhanced star formation in these objects. Low mass haloes are thus more likely to host stars (or a galaxy) in \fr which causes the stars to be less clustered compared to standard cosmology. 
\item The halo mass function is enhanced in the considered modified gravity models compared to GR. The enhancement depends on halo mass and redshift. The strongest effect can be observed in the mass regime in which haloes became unscreened recently. This \fr effect is consistent with what was found in previous DM-only simulations \citep{winther2015, schmidt2010, arnold2018}. 
The baryonic feedback suppresses the formation of haloes around $m_{200}^{\mathrm{crit}} \approx 10^{11} h^{-1}M_\odot$ relative to DM-only. The suppression is independent of the cosmological model.
One can thus conclude that the effects of baryons and modified gravity on the halo mass function can be treated independently in separate simulations. 
\item The galaxy host halo mass function is unaffected by \fr at high masses (where all haloes host central galaxies). In the transition region, where only part of the haloes are occupied by a galaxy, the host halo mass function is significantly enhanced in \fr compared to GR due to the enhanced star formation rates in modified gravity.
\item The number of sub-haloes is not significantly affected by \fr. This is true for both the total subhalo number identified by \textsc{subfind} and for the number of luminous subhaloes (satellite galaxies) per host halo.
\item The halo-halo auto-correlation function is suppressed in \fr. The results are consistent with \cite{arnold2018}. The baryonic effects do not show a strong dependence on the gravity model. 
\end{itemize}

We studied the matter statistics in \fr in the SHIBONE simulations. While the simulations for the first time allow to study the combined effect of chameleon type modified gravity and baryonic feedback from stars and AGN we find that the uncertainties related to baryonic processes still leave many open questions for the future. This is particularly true if percent-level accurate fitting formulae which can be used to analyse upcoming large scale structure surveys are desired. 


\section*{Acknowledgements}
The authors like to thank Volker Springel, Rainer Weinberger and R\"udiger Pakmor for their help with the \textsc{arepo} code and the simulations and for useful comments on the results. We also thank the IllustrisTNG collaboration for allowing us to use their galaxy formation model. The authors are furthermore grateful to Jian-hua He and Carlos Frenk for insightful comments and discussions. 

CA and BL acknowledge support by the European Research Council through grant ERC-StG-716532-PUNCA. BL is also supported by the STFC through Consolidated Grants ST/P000541/1 and ST/L00075X/1.

This work used the DiRAC@Durham facility managed by the Institute for
Computational Cosmology on behalf of the STFC DiRAC HPC Facility
(www.dirac.ac.uk). The equipment was funded by BEIS capital funding
via STFC capital grants ST/K00042X/1, ST/P002293/1, ST/R002371/1 and
ST/S002502/1, Durham University and STFC operations grant
ST/R000832/1. DiRAC is part of the National e-Infrastructure.

\bibliographystyle{mnras}
\bibliography{paper}


\end{document}